\documentclass[a4paper,10pt,twoside]{cpc-hepnp}

\usepackage{multicol}
\usepackage{graphicx}
\usepackage{booktabs}
\usepackage{amssymb,bm,mathrsfs,bbm,amscd}
\usepackage[tbtags]{amsmath}
\usepackage{lastpage}

\begin{document}

\fancyhead[c]{\small Submitted to Chinese Physics C}
\fancyfoot[C]{\small }

\footnotetext[0]{}

\title{The Induced Charge Signal of Glass RPC Detector\thanks{Supported by the Fundamental Research Funds for the Central Universities£¬Southwest University for Nationalities (JB2012092) }}

\author{%
  HAN Ran$^{1),2)}$\email{han.ran@gmail.com}%
}
\maketitle

\address{%
$^1$ School of Nuclear Science and Engineering, North China Electric Power University,  Beijing,  102206,  China\\
$^2$ The Institute of Nuclear physics of Lyon,Villeurbanne,69622,France
}

\begin{abstract}
The gas detector glass resistivity parallel chamber (GRPC) is proposed to use in the Hadron calorimeter (HCAL), the read-out system is based on the semi-digital system, then the charge information from GRPC is needed. To better understand the charge comes out from GRPC, we started from cosmic ray test to get the charge distribution and then study the induced charge distribution on the collection pad, after successfully compared with the prototype beam test data at CERN (European Council for Nuclear Research), the process was finally implanted into the Geant4 based simulation for future study.
\end{abstract}



\begin{multicols}{2}

\section{Introduction}

The International Large Detector (ILD) is a concept for a detector at the International Linear Collider (ILC). The ILC will collide electrons and positrons at energies of initially 500GeV, upgradeable to 1 TeV\cite{lab1,lab2}. The ILC has an ambitious physics program, which will extend and complement that of the Large Hadron Collider (LHC). The design of ILD, is driven by these requirements. Excellent calorimetry and tracking are combined to obtain the best possible overall event reconstruction, including the capability to reconstruct individual particles within jets for particle flow calorimetry. Within the ILD paradigm of particle flow calorimetry\cite{lab3},the ultimate jet energy resolution is achieved by reconstructing charged particles in the tracker, photons in the electromagnetic calorimeter (ECAL), and neutral hadrons in the ECAL and hadronic calorimeter (HCAL).

The capacity to apply successfully the particle flow algorithms can be enhanced by increasing the granularity of the different ILD sub-detectors. In the hadronic calorimeter this will doubtlessly help reduce the confusion between charged and neutral hadronic particles by providing a better separation of the associated showers. However, the cost related to such an increase in detector segmentation should be minimized. To satisfy both, a gas hadronic calorimeter with a semi-digital readout is proposed. The choice of gaseous detectors as the sensitive medium in the HCAL offers the possibility to have very fine segmentation while providing high detection efficiency.

\section{The Structure of GRPC}

The (GRPC) is one of these gaseous detectors proposed to be used in ILD, which can be built in large quantities at low cost. Large GRPCs as the ones required for the ILD HCAL can be easily produced\cite{lab4}. This is an important advantage with respect to other detectors since it guarantees very good homogeneity. However, the GRPCs to be used in the ILD HCAL need to be more elaborate. As the HCAL is situated inside the magnet coil, the sensitive medium thickness is an important issue. Very thin GRPCs are requested and 3.0mm thick GRPCs were indeed produced and successfully tested. In Fig.~\ref{fig1} a scheme of such a single gap GRPC is shown.
\begin{center}
\includegraphics[width=9cm]{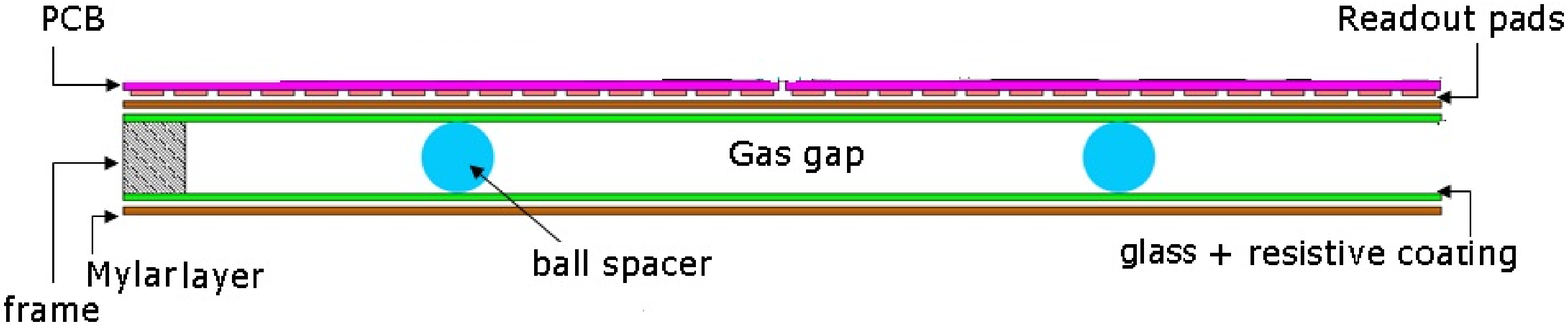}
\figcaption{\label{fig1}A scheme of single gap Glass RPC.  }
\end{center}

To better understand the performance of GRPC, the prototype was simulated and validates by beam test data. In the simulated prototype we have 40 layers, in each layer there is 2cm steel absorber. The size of each layer is 1m*1m, which is identical to prototype. In addition to the absorber each layer has one GRPC chamber which is composed by one gas gap, two glass plates, and the outer side of the glass plates was covered by a thin layer of resistive coating. A Mylar layer of 50 microns separates the anode from the pads of the electronic board which has the same size of the GRPC; the pad is 1cm by 1cm, so we totally have 9216 readout channels in simulation.

\section{The induced signal of GRPC }

The simulation is based on Geant4, the output from Geant4 is the deposited energy, in order to understand GRPC performance and achieve better comparison with data, the induced charge distribution is also studied and implanted into Geant4 package.
Many papers have described the physics process of avalanche growth and induced charge\cite{lab5,lab6}. Here we are using one of them to describe the induced charge distribution and to compare it with test beam data.
The induced charge spectrum of RPC in avalanche mode can be described by a Polya function \cite{lab7}:                                                                                                 \begin{equation}
\label{eq1}
Q=c\frac{(b+1)^{b+1}}{b!}(\frac{x}{a})^{b}e^{(-(b+1)\frac{x}{a})}
\end{equation}

 $a=e^{\alpha s}$is average multiplication factor on the gap $s$ ,$b$ is an integer to determine the shape and $c$ is a normalization factor.
 From Eq.~(\ref{eq1}), we would expect that if the parameters are set properly, the charge distribution in simulation will be well described. In this case we fixed those parameters using the cosmic ray test data.

\subsection{The charge spectrum of cosmic ray}

The test setup is shown in Fig.~\ref{fig2}. The detector is made of a small GRPC (32X8cm2) equipped with 64-pads electronic board can be readout individually using an oscilloscope connected to a PC on which a Labview-based DAQ system was used to analysis the analog output signal. The trigger system is made of two scintillators with an overlapping area smaller than that of the GRPC. The avalanche signal charge spectrum that was collected from few thousands of events is shown in Fig.3, together with its Polya distribution fitting curve. The parameters from Polya function $p0,p1,p2$ are also shown in the Fig.~\ref{fig3}.
\begin{center}
\includegraphics[width=6.5cm]{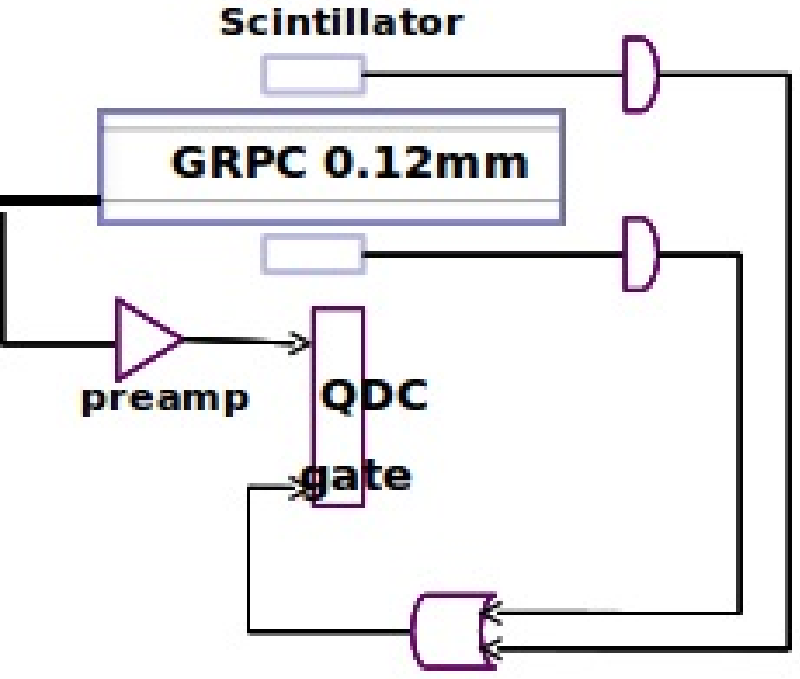}
\figcaption{\label{fig2}  Cosmic charge spectrum test setup. }
\end{center}
\begin{center}
\includegraphics[width=7cm]{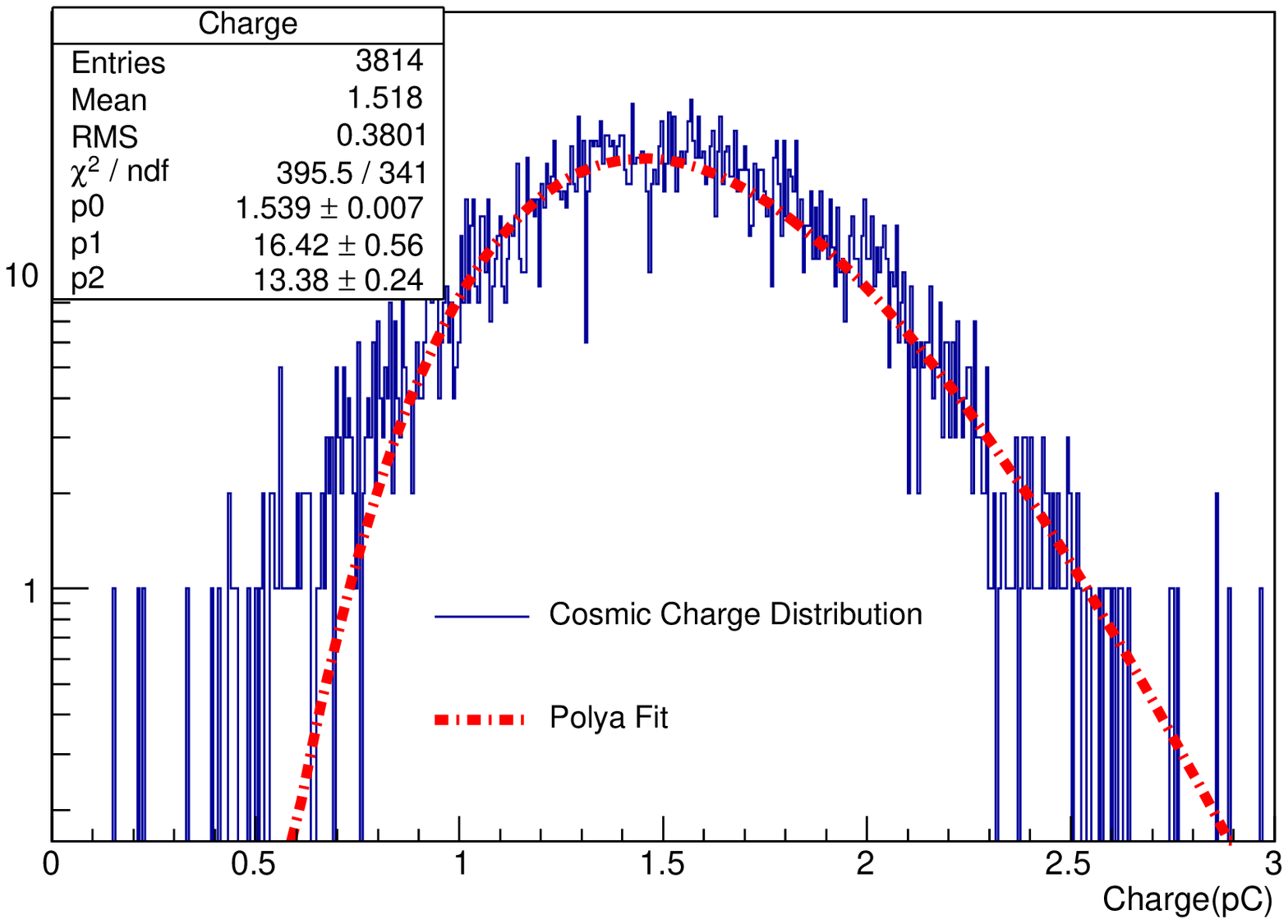}
\figcaption{\label{fig3}  Typical avalanche signal charge spectrum and its Polya fitting curve. }
\end{center}
To get the charge distribution on pickup pad, two effects are included in the study; 1) Impact spatial coordinates $x \& y$, 2) Smearing due to resistive coating.
\subsection{The induced charge distribution on pickup pad}

To calculate the induced charge distribution on the pad plane, we considered a simplified model as shown in Fig.~\ref{fig4}. Here, $a$ is the gas gap, $q$ is the total charge getting from polya function (the charge we got after avalanche), we take it as a point charge and $d $ is the location of $q$.

\begin{center}
\includegraphics[width=6.5cm]{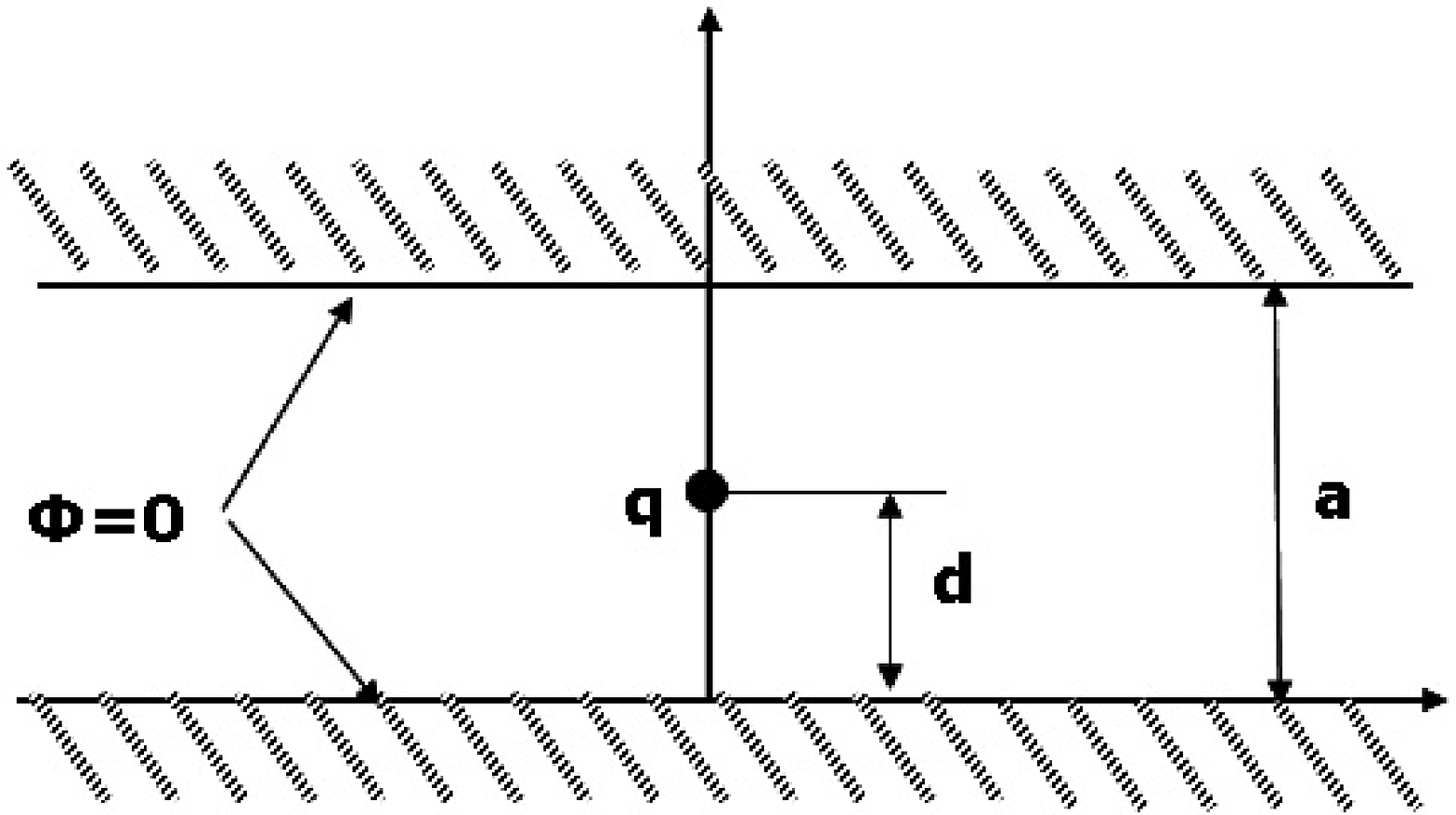}
\figcaption{\label{fig4}  Model to calculate the induced signal distribution on the pink up pad. }
\end{center}
The potential in the gap can be expressed as following equation \cite{lab8},

\begin{equation}
\label{eq2}
\Phi(x,y)=4q\Sigma\frac{1}{n}\sin(\frac{n\pi d}{a})\sin(\frac{n\pi y}{a})\exp(\mp(\frac{n\pi x}{a}))
\end{equation}
if deriving the induced charge on the surface $y=0$, then we obtain the induced charge density distribution:
\begin{equation}
\label{eq3}
\sigma(x)=\frac{-q}{2a}\frac{\sin(\pi \frac{d}{a})}{\cosh(\pi \frac{x}{a})-\cos(\pi \frac{d}{a})}
\end{equation}

There is no big difference for the charge $q$ at the gap center, bottom or top, so we take it at center which is $a=2d$. Then Eq.~(\ref{eq3}) will be like
\begin{equation}
\label{eq4}
\sigma(x)=c\frac{-q}{2a}\frac{1}{cosh(\pi\frac{x}{a})}
\end{equation}

In Eq.~(\ref{eq4}), $c$ is normalize factor which makes  $\int_{-\infty}^{\infty}\sigma(x)dx=q$,$ a=0.12cm$ for GRPC.
The above equation corresponds to the case of $y=0$, but for in our pad case which is two dimensional, so we rewrite charge density distribution
\begin{equation}
\label{eq5}
\sigma(x,y)=c\frac{-q}{2a}\frac{1}{\cosh(\pi\frac{\sqrt{{(x-x_0)^2+(y-y_0)^2}}}{a})}
\end{equation}
$x_0$,$y_0$ are the position of $q$ , the two dimensional induced charge distribution is plot in Fig.~\ref{fig5}.
\begin{center}
\includegraphics[width=9cm]{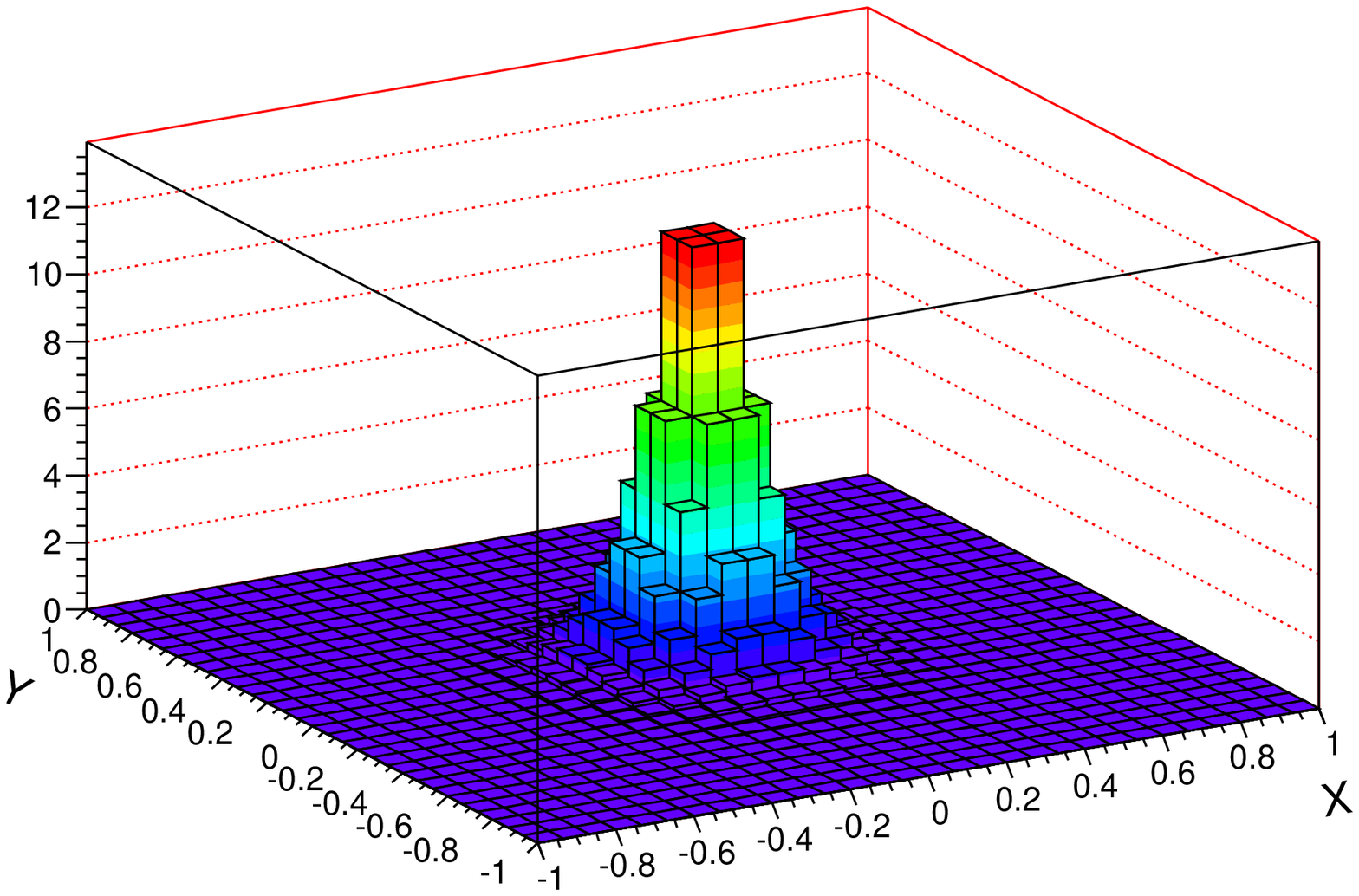}
\figcaption{\label{fig5} The distribution of two dimensional induced charge on pad. }
\end{center}

\subsection{Painting effect}
If only spatial effect is considered, the cluster size of GRPC pad readout should be 1 or 2, but in reality, this is not always true. One additional consideration is that the induced charge distribution is smeared by graphite coating. Between RPC gap and the pickup pad plane there is a graphite coating layer. Depending upon the value of the resistivity of this coating layer, the distribution of the induced charge on the pickup plane the smearing effect can be more or less large. In the simulation we consider this affect by increase the distance of gas gap.

\section{The efficiency and multiplicity comparison with data}
We first compared the efficiency and multiplicity with beam test at CERN 2009, the beam is 7GeV$\pi+$, the gas gap plus graphite effect lead to gas gap $a=0.24cm$. Fig.6 show the efficiency comparison between beam test results and simulation results, dot points are data from 2009 beam test at CERN, inverted triangle for simulation results. From the plot we can clear see that with different threshold, the data and simulation are in good coincidence, Fig7 is the same but for multiplicity comparison, dot points are data from 2009 beam test at CERN, triangle for simulation results.

\begin{center}
\includegraphics[width=7cm]{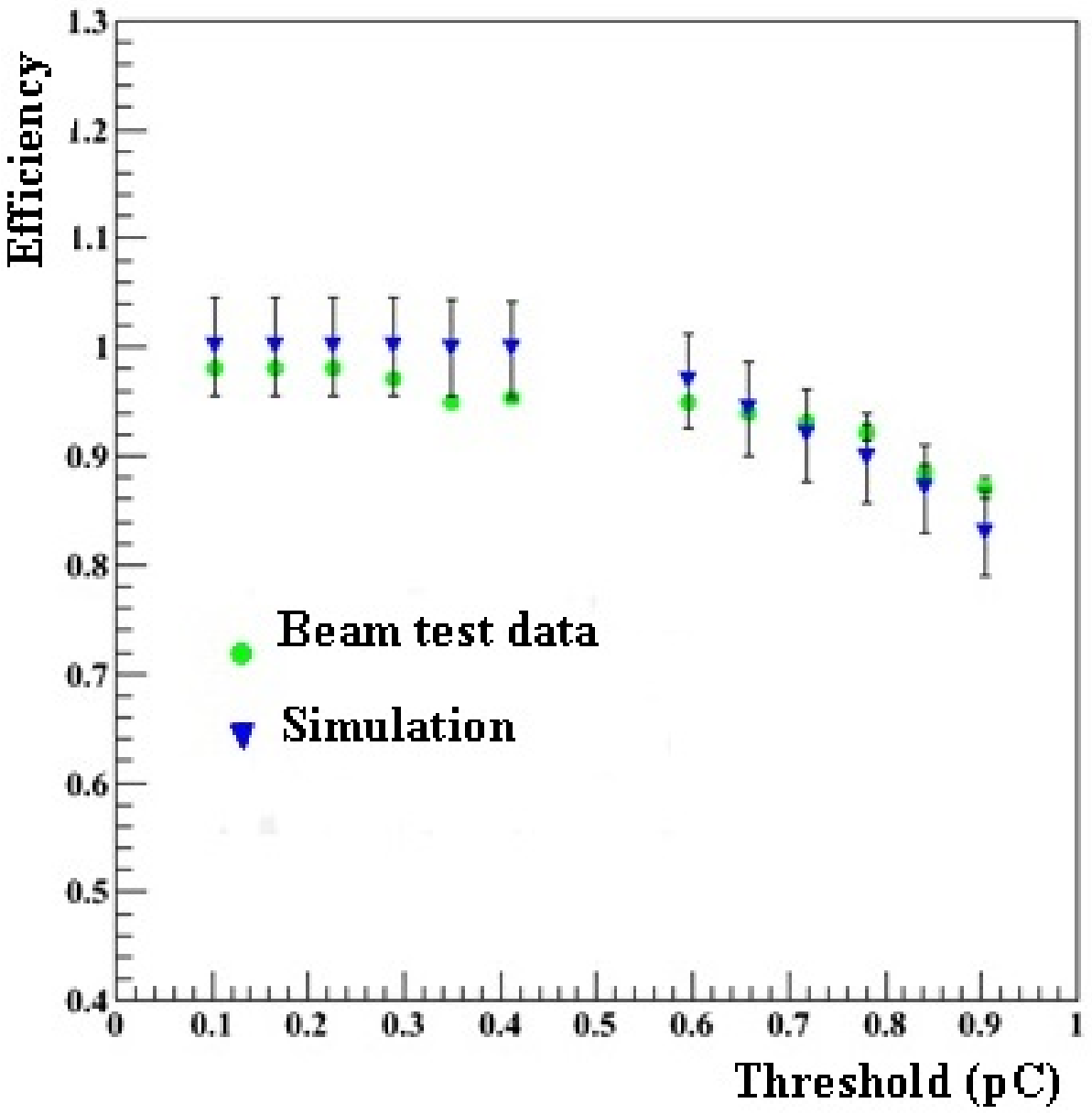}
\figcaption{\label{fig6}  Efficiency $.vs.$ threshold at a=0.24cm in simulation. }
\end{center}
\begin{center}
\includegraphics[width=6.5cm]{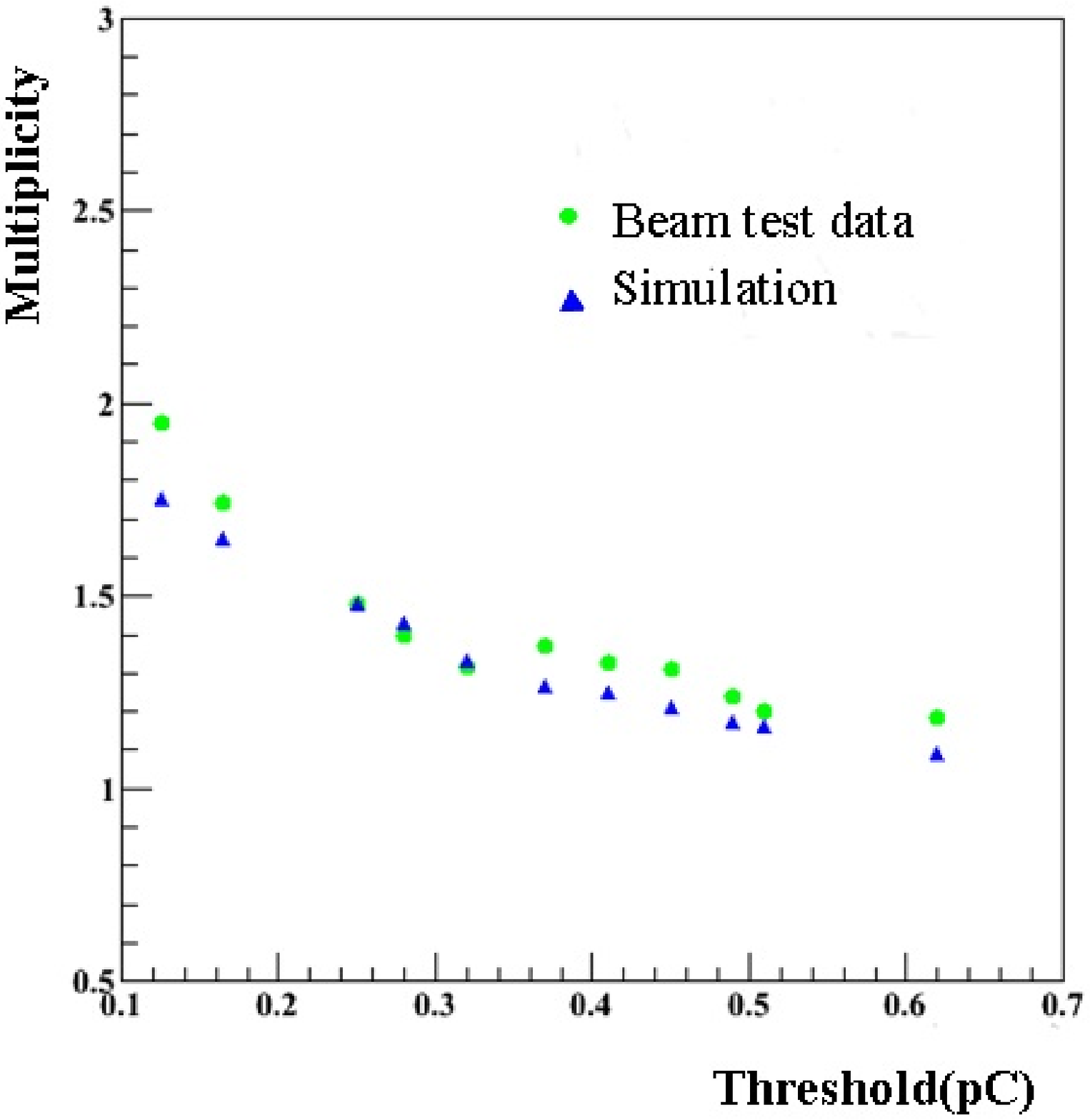}
\figcaption{\label{fig7}  Pad multicity $.vs.$ threshold.}
\end{center}

\section{conclusion}
A simulation model based on MOKKA-GEANT4\cite{lab9} package was developed aiming at reproducing the detector response as observed in real data. The model provides output format identical to the one to be used for data so future comparison between data and simulation can be straightforward.
The model will be improved be more data expected from coming Test Beam where few units will be exposed to pions and muons beams of different energies.

\end{multicols}

\vspace{15mm}

\vspace{-1mm}
\centerline{\rule{80mm}{0.1pt}}
\vspace{2mm}

\begin{multicols}{2}

\end{multicols}

\clearpage

\end{document}